\documentclass{optica-article}
\journal{opticajournal} 
\articletype{Research Article}
\usepackage{lineno}
\usepackage{hyperref}
\hypersetup{
    colorlinks=true,
    linkcolor=blue,
    anchorcolor=blue,
    citecolor=blue,
    urlcolor=blue
}

\begin{document}

\title{Reconstruction of partially occluded objects with a physics-driven self-training neural network}

\author{Mingjun Xiang\authormark{1,2,3}, Kai Zhou\authormark{4,1,3}, Hui Yuan\authormark{2,*}, and Hartmut G. Roskos\authormark{2}}

\address{\authormark{1}Frankfurt Institute for Advanced Studies (FIAS), 60438 Frankfurt am Main, Germany\\
\authormark{2}Physikalisches Institut, Goethe-Universität Frankfurt am Main, 60438 Frankfurt am Main, Germany\\
\authormark{3}Xidian-FIAS International Joint Research Center, 60438 Frankfurt am Main, Germany\\
\authormark{4}School of Science and Engineering, The Chinese University of Hong Kong, Shenzhen, P.R. China}

\email{\authormark{*}yuan@physik.uni-frankfurt.de} 

\begin{abstract*} 
This study proposes a novel approach utilizing a physics-informed deep learning (DL) algorithm to reconstruct occluded objects in a terahertz (THz) holographic system. Taking the angular spectrum theory as prior knowledge, we generate a dataset consisting of a series of diffraction patterns that contain information about the objects. This dataset, combined with unlabeled data measured from experiments, are used for the self-training of a physics-informed neural network (NN).
During the training process, the neural network iteratively predicts the outcomes of the unlabeled data and reincorporates these results back into the training set. This recursive strategy not only reduces noise but also minimizes mutual interference during object reconstruction, demonstrating its effectiveness even in data-scarce situations.
The method has been validated with both simulated and experimental data, showcasing its significant potential to advance the field of terahertz three-dimensional (3D) imaging. Additionally, it sets a new benchmark for rapid, reference-free, and cost-effective power detection.
\end{abstract*}

\section{Introduction}
THz radiation encompasses electromagnetic waves within the frequency range of 0.1~THz to 10~THz, corresponding to wavelengths between 3~mm and 30~$\mu$m. As a pivotal area of THz research, THz holographic imaging technology \cite{Heimbeck:20, s21124092} leverages the distinctive physical properties of this frequency band. Firstly, THz waves have strong penetration capabilities, enabling them to pass through non-metallic and non-polar materials such as ceramics, plastics, and foam, which are typically opaque to visible and infrared light. Secondly, the low photon energy of THz radiation prevents harmful ionization effects. Thirdly, numerous molecules exhibit unique absorption and dispersion characteristics within the THz frequency range, allowing the creation of molecular fingerprint spectra for substance identification \cite{kawase2003non}. Furthermore, THz waves are highly sensitive to water content, making them ideal for assessing material hydration levels \cite{thrane1995thz}. Over the last two decades, progress in electronic and photonic technologies has significantly advanced THz imaging technology \cite{6005341}. This progress has broadened its application to areas including non-destructive testing \cite{Hasegawa2003}, quality control \cite{gowen2012terahertz}, security screening\cite{Hasegawa2003}, and biomedical imaging \cite{pickwell2006biomedical}. These advancements have resulted in significant successes, showcasing advantages that are not achievable with holographic imaging in other frequency bands.

Occlusion is one of the critical challenges that urgently needs to be addressed in imaging technology. The occluded parts of objects often cannot be directly imaged, affecting the completeness and accuracy of the image. This issue is particularly pronounced with metallic objects, as terahertz waves cannot penetrate metal, rendering traditional back-propagation reconstruction methods ineffective in eliminating mutual interference between two occluded objects \cite{Yuan19,yuan2023a}. For the second object, the beam is no longer Gaussian due to interactions with the first object, and significant backscattering and multiple diffraction may occur. To tackle this issue, researchers have proposed various methods to address occlusion problems: 
Using binocular cameras or detector arrays to capture images from multiple angles \cite{geiger1995occlusions,ren2021effective}, acquiring data from different perspectives, and improving the reconstruction of occluded areas through multi-view fusion techniques. However, these methods are mostly applied in the visible light spectrum, while terahertz hardware has not kept pace. 
Another approach is integral imaging \cite{Shin:08}, which reconstructs 3D structures from multi-view images and estimates occluded areas using depth maps. Although integral imaging provides detailed 3D information, processing a large number of view images requires high computational costs, and obtaining high-resolution images in the terahertz domain is also a significant challenge. 
The optical flow method employs optical flow estimation and sub-pixel precision depth extraction techniques to reconstruct occluded 3D objects \cite{Jung:10}, but this method also requires high-resolution images. These methods demonstrate varying degrees of potential in addressing occlusion problems but still face numerous technical bottlenecks in the THz domain.

Image reconstruction can be considered an inverse problem \cite{xiang2024amplitude}. Recently, DL techniques have emerged as a promising and effective approach for solving inverse problems \cite{Pang:2016vdc,Shi:2021qri,Wang:2021jou}, primarily relying on supervised deep learning methods using labeled experimental data. However, the time-consuming acquisition of terahertz images poses challenges for supervised training using experimental data.

In this paper, we propose an innovative method for reconstructing occluded objects in THz holography using a physics-driven self-training neural network. Our approach involves training the network with only the visible light MNIST dataset \cite{cohen2017emnist} and a limited amount of experimental data. By integrating the angular spectrum theory and adhering to physical principles, we generated 30,000 datasets for supervised pre-training, forming a network capable of accurately predicting simulated data. Subsequently, we employed experimental data for adaptive training, continuously optimizing the prediction results to ultimately retrieve information about the occluded objects. The model demonstrated rapid prediction speed and noise robustness in experimental data, and accurately predicted alphabetic objects as well. Our findings highlight the potential of reconstructing occluded objects from THz diffraction patterns.

\section{Method}
\begin{figure}[ht]
\centering
\includegraphics[width=\linewidth]{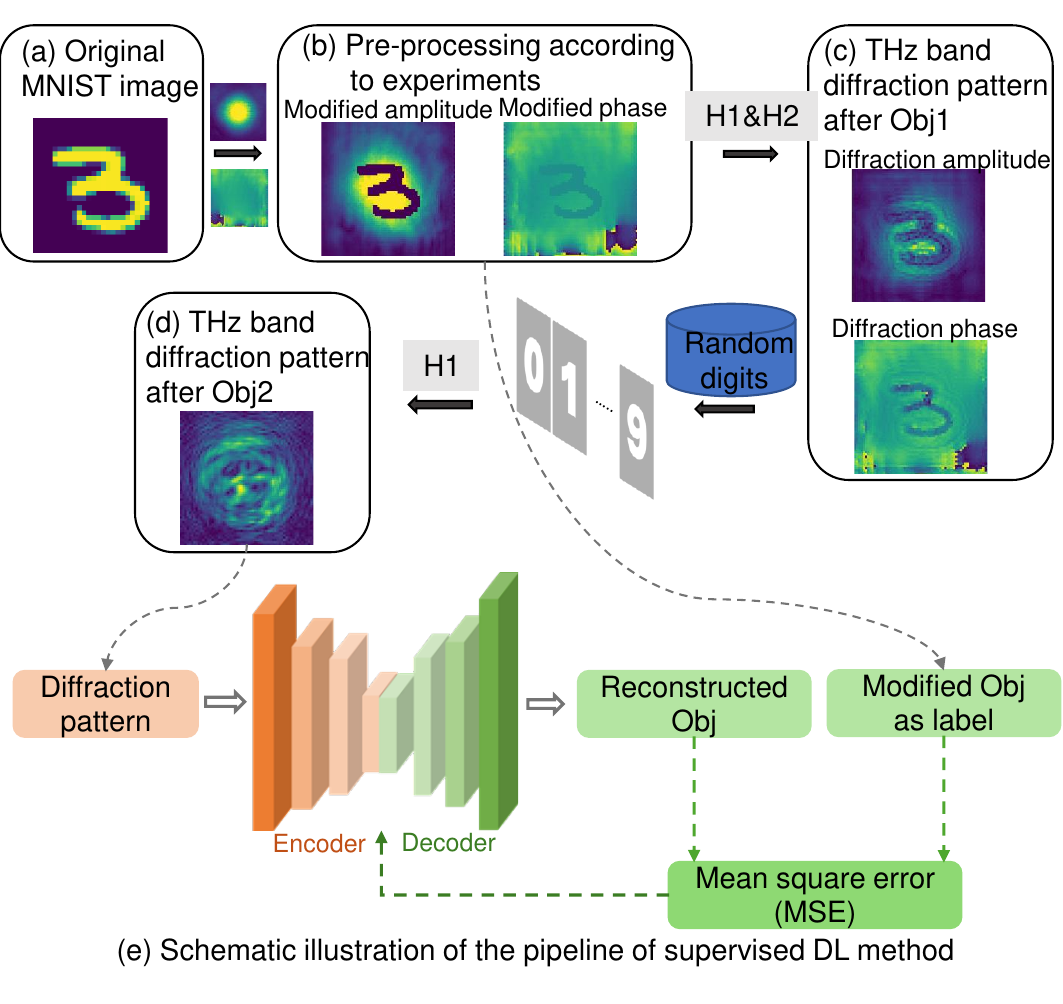}
\caption{Illustration of the dataset preparation and NN pre-training. }
\label{fig:false-color1}
\end{figure}
Figure~\ref{fig:false-color1} illustrates the process of preparing labeled data and my pre-training supervised model. 
In particular, Fig.~\ref{fig:false-color1}(a) showcases randomly selected original images sourced from the MNIST dataset. Preprocessing begins by superimposing the measured Gaussian field amplitude and phase distribution of the THz beam onto these images, thereby yielding modified amplitude and phase representations influenced by the material properties and thickness of the object, as depicted in Fig.~\ref{fig:false-color1}(b). Subsequently, leveraging physical models $H_1$ and $H_2$ (outlined below), we calculate the diffraction amplitude and phase of the THz image using the angular spectrum approach \cite{goodman2005introduction}, as illustrated in Fig.~\ref{fig:false-color1}(c). This diffraction field serves as the input field for the second object, selected from random digits, and is processed anew using the physical model. Then, a series of corresponding diffraction patterns and object information is generated.
This pattern is fed as input into the convolutional neural network (CNN) as shown in Fig.~\ref{fig:false-color1}(e). CNNs are widely used in image reconstruction tasks because of their ability to efficiently capture local features through shared weights and local receptive fields, their robustness to translation invariance\cite{Zhou:2018ill}, and their hierarchical feature extraction capability\cite{alzubaidi2021review}. Its architecture is inspired by U-Net \cite{alzubaidi2021review} and consists of a down-sampling path as well as a symmetric up-sampling path (see \nameref{AppendixA} for a detailed description). The output of the CNN is reconstructed first object images. The mean squared error (MSE) between this image reconstruction and the ground truth image (i.e., the modified amplitude image in (b)) is taken as the loss function to optimize the CNN. In image processing, MSE is commonly used to measure the quality of the reconstructed image compared to the ground truth. It represents the Euclidean distance between images and is simple and efficient to evaluate, making it suitable for large-scale image processing tasks. Additionally, MSE is a convex function (in terms of the network output) and possesses good mathematical properties \cite{4775883}. In a subsequent step, the same network is applied for the second object retrieval from the diffraction patterns.

As mentioned, the physical model $H_1$\&$H_2$ simulates the THz imaging process in the experiment. If a planar object is illuminated by a collimated continuous-wave beam with a given cross-sectional amplitude profile and a zero phase offset, the complex-valued field amplitude right behind the object can be written as 
\begin{equation}
E(x,y,z=0) = A(x,y,0) e^{i\phi(x,y,0)} ,
\label{eq:refname1}
\end{equation}
where $A(x,y,0)$ and $\phi(x,y,0)$ are the amplitude and the phase of the transmitted beam. Over a distance $d$, diffraction reshapes the field to \cite{goodman2005introduction}
\begin{equation}
E(x,y,z=d) = \iint \hat{E}_0(f_x,f_y)\,G\,e^{i2\pi(f_x x+f_y y)} df_x\,df_y
\label{eq:refname2}
\end{equation}
with $G=e^{ikd\sqrt{1-\lambda^2f_x^2-\lambda^2f_y^2}}$ the wave propagation function, and $\lambda$ the wavelength, $\hat{E}_0$ the spatial Fourier transform of $E(x,y,z=0)$ with $f_x=x/(\lambda d)$ and $f_y=y/(\lambda d)$ as the spatial frequencies in the $x$ and $y$ directions. The diffraction pattern is the absolute value of the propagated field
\begin{equation}
A(x,y,z=d) = \left|E(x,y,z=d)\right| = H_1(\phi(x,y,0), A(x,y,0), d),
\label{eq:refname3}
\end{equation}
the diffracted phase is the field's phase angle value
\begin{equation}
\phi(x,y,z=d) = Im(ln(E_d(x,y,z=d))) = H_2(\phi(x,y,0), A(x,y,0), d),
\label{eq:refname4}
\end{equation}
where $H_1$ and $H_2$ represent the mapping function relating the object's amplitude and phase information to the diffraction amplitude and phase. 

For the second object, the diffraction process is similar to that of the first object. The only difference is that, in the final analysis, we only need to obtain the diffraction pattern, specifically the diffraction amplitude. This is because the diffraction amplitude is the data we can acquire experimentally and is the sole input we intend to use for reconstructing the two objects.

Once sufficient data has been collected, we need to train a CNN to fit the relationship between the diffraction patterns and the first object (as well as the second object, using two different networks). Specifically, we will construct two distinct NNs: one to fit the relationship between the diffraction pattern and the first object, and another to fit the relationship between the diffraction pattern and the second object. The core of this approach lies in leveraging the powerful nonlinear fitting capabilities of neural networks, learning the complex mapping between the diffraction patterns and the structures of the objects through extensive training data.

The relation function $R_\theta$ ($\theta$ denotes the network weights and bias parameters with $\theta^*$ is their optimum) from a large number of labeled data $(A(x,y,z=d)_k, d_k)$ contained in a labeled training set $S_T = {(A(x,y,z=d)_k, d_k), k=1,2,..., K}$, with its optimum defined as:
\begin{equation}
\theta^* = \mathop{\arg\min}_{\theta} \|R_\theta(A(x,y,z=d)_k)-A(x,y,0)\|^2 \qquad\forall(d_k,A_k)\in S_T.
\label{eq:refname5}
\end{equation}

\begin{figure}[ht]
\centering
\includegraphics[width=\linewidth]{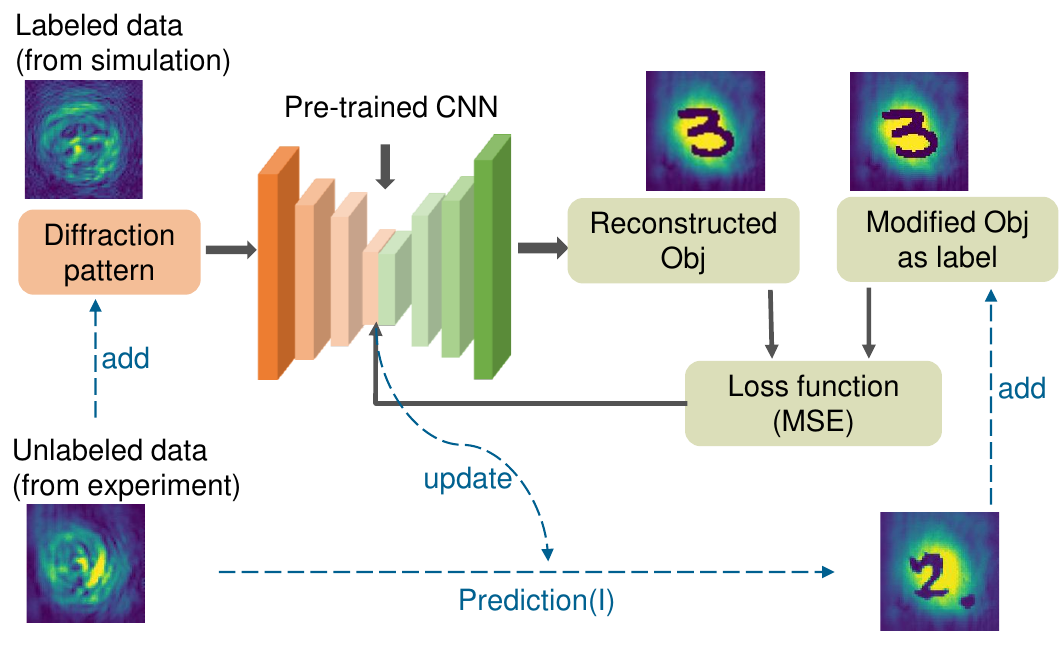}
\caption{Schematic pipeline illustration of the proposed object reconstruction algorithm. }
\label{fig:false-color2}
\end{figure}
Upon completing the pre-training of the network, we save all the network parameters to use as initial values for the subsequent training phase. As illustrated in Figure~\ref{fig:false-color2}, the network's input remains the diffraction patterns. However, this now includes both labeled data from simulations and unlabeled data from experiments. The labeled simulation data can be several thousand sets; for instance, in our case, we used 1,000 sets, distinct from the data used in pre-training the network. The experimental unlabeled data consists of 22 sets and we expanded it to 66 sets with different noise in our example.
For the labeled data, we directly compute the MSE between the network’s reconstruction and the labels. For the unlabeled data, we first use the network parameters stored from the previous epoch to make predictions, then set these predictions as the labels. We then calculate the MSE between the network's predictions and these labels for the current epoch. This cumulative error is used to adjust the network parameters during the training iterations.
As the network iterates, its predictions for the unlabeled data gradually stabilize. It is worth noting that due to the challenges and lengthy time required to collect data in the THz frequency range, each 80*80 data set takes approximately one hour to gather. Despite the limited amount of experimental data, our method demonstrates both accuracy and robustness.

\section{Results}
In the following, we will demonstrate the performance of the proposed method
using data derived from simulations (THz-emulated MNIST) and experiments, respectively.
\begin{figure}[ht]
\centering
\includegraphics[width=\linewidth]{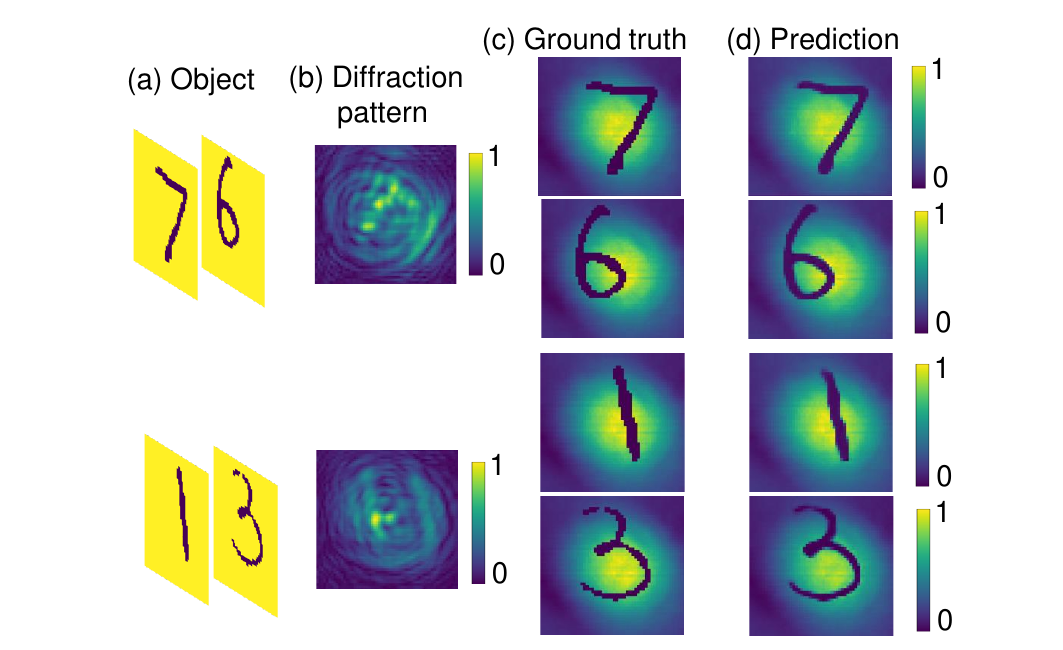}
\caption{Reconstruction results for the simulation data; Upper half from the existing training dataset, lower half from the blind data for the NN.}
\label{fig:false-color3}
\end{figure}

Upon completing the pre-training phase, we utilized an additional 1,000 sets of labeled data, combined with augmented experimental data, to perform self-training until achieving a well-fitted network. The upper part of Figure~\ref{fig:false-color3} shows the results for one of these 1,000 datasets previously seen by the network, while the lower part presents the results for an entirely new dataset never encountered during training.
In the example shown in the upper part, the combination consists of the digits seven and six. The final diffraction pattern of this combination is depicted in Fig.~\ref{fig:false-color3}(b). Using this diffraction pattern as the sole input for network prediction, the reconstruction result is shown in Fig.~\ref{fig:false-color3}(d). The computed MSE for the prediction of the second object (digit '6') is only 0.0035, while the MSE for the first object (digit '7') is 0.0038. The reason the MSE for the second object is lower than that of the first object is due to its closer proximity to the detector. Once the prediction network was thoroughly trained, it consistently showed that the overall loss for the second object was slightly smaller compared to the first object.
The lower part of the Figure~\ref{fig:false-color3} features a combination of the digits one and three, which were never present in the training dataset. Despite this, the network's predictions remain remarkably accurate, with the MSE for the second object (digit '3') being only 0.0041 and for the first object (digit '1') being 0.0043. This demonstrates that the network can provide accurate predictions even when faced with novel data combinations, thus validating its generalization capability on unseen data.

\begin{figure}[ht]
\centering
\includegraphics[width=\linewidth]{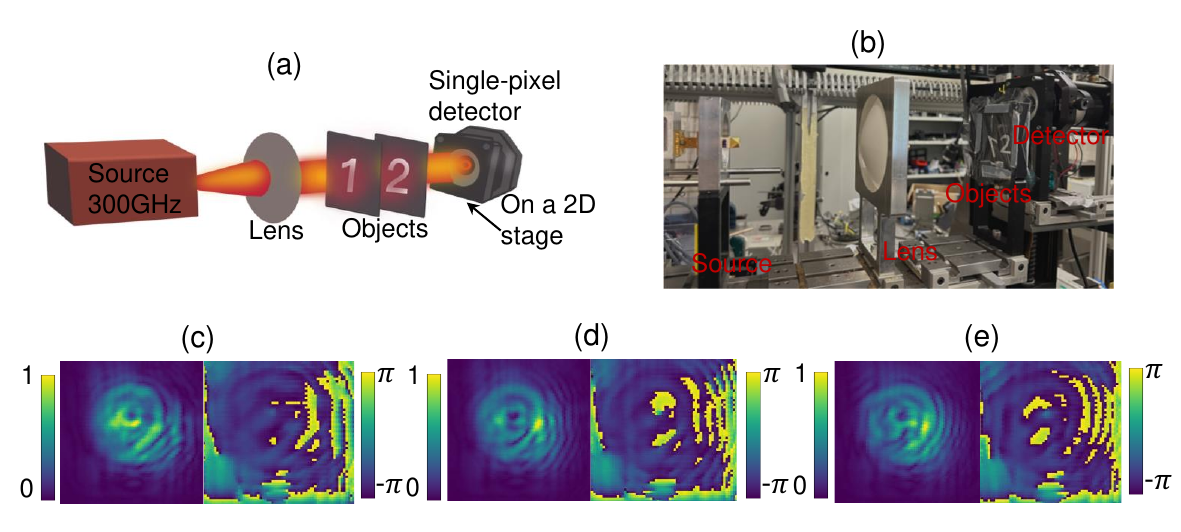}
\caption{Experimental setup and results. (a, b) Schematic and photograph of the THz in-line-holographic imaging system (LO-wave injection not shown), (c-e) measured diffraction amplitudes and phases.}
\label{fig:false-color4}
\end{figure}

Next, I will elaborate on the process of experimental data collection. Initially, we utilized heterodyne detection system to acquire phase information and employed back-propagation for reconstruction. This approach is highly effective and serves as a benchmark for comparison in our research. It is noteworthy that in the comparative results, phase information was employed. Conversely, our proposed method leverages a self-training neural network that exclusively utilizes the intensity information of the object, specifically the amplitude information, as depicted in the left half of Figure~\ref{fig:false-color4} (c-e). Consequently, this method obviates the need for local oscillator signal for reconstruction.

The schematic diagram and a photograph of our experimental setup is illustrated in Figure~\ref{fig:false-color4} (a) and (b), respectively. In our experiments, we employed a 300 GHz frequency multiplier chain (vendor: Virginia Diodes, Inc.; output power: approximately 1 mW), which was collimated by a focusing lens with a focal length of 10~cm and an aperture of 4~inches. To satisfy the requirements for a large numerical aperture, we selected an aspherical hyperbolic lens to mitigate spherical aberration, a common issue with spherical lenses. The aperture of lens needed to be sufficiently large to ensure that the object was illuminated by a collimated Gaussian beam, thereby preventing beam amplitude modulation caused by diffraction at the lens edges. This lens was fabricated from polytetrafluoroethylene (PTFE) to ensure precision and stability.
In terms of detection, we employed a single-pixel TeraFET detector \cite{Ikamas2018}, which was mounted on a high-precision two-dimensional translation stage. This translation stage was meticulously controlled to scan and record the diffraction patterns point by point. The radiation from the object entered the detector chip from the backside and traversed a 4~mm diameter silicon substrate lens \cite{PIERS19, IWMTS22}. This configuration enabled the detector to efficiently capture the weak signals reflected or transmitted by the object, while circumventing power loss and signal attenuation issues that might be encountered with conventional detectors.

In our experiment, we used two metal objects placed at distances of \(d = 75\)~mm and \(d = 90\)~mm from the detector. These objects were made of aluminum foil with a thickness of approximately 15 micrometers and lateral dimensions of approximately \(15 \times 25\)~mm², adhered to a 10-micrometer-thick polypropylene film. The scanning image area was \(80 \times 80\) pixels with a pixel size of 1~mm². We incrementally moved the objects to increase their overlapping regions, and recorded the resulting diffraction patterns and phase changes, as shown in Figure~\ref{fig:false-color4} (c-e). The left side of the figure displays the amplitude, while the right side shows the phase.

\begin{figure}[ht]
\centering
\includegraphics[width=\linewidth]{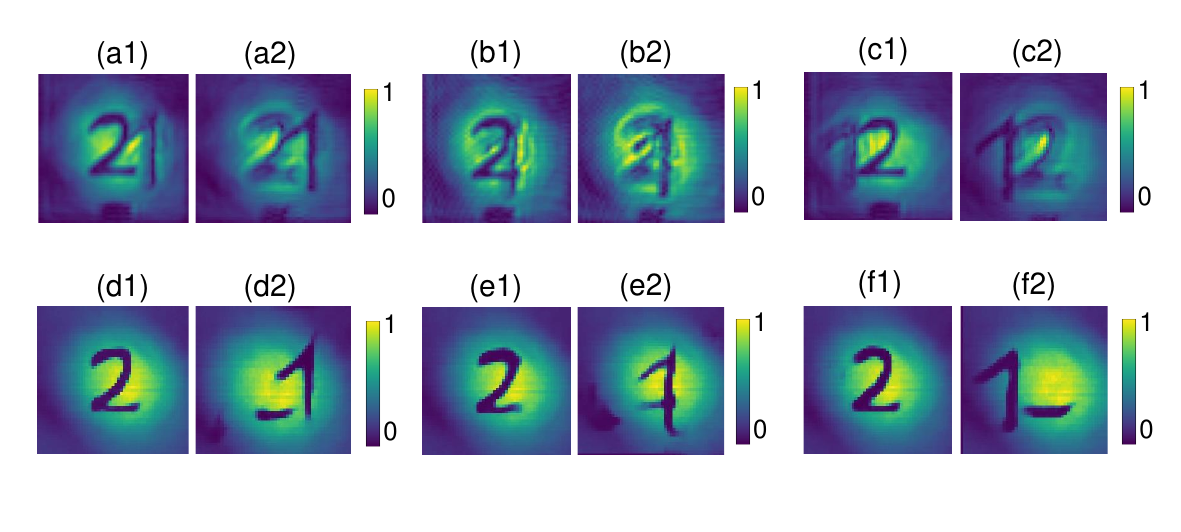}
\caption{Reconstruction results of digit objects using the back-propagation method and our proposed self-training neural network.}
\label{fig:false-color5}
\end{figure}

We first used the amplitude and phase data obtained from heterodyne detection (as shown in Fig.~\ref{fig:false-color4} (c-e)) for back-propagation. The reconstruction results at distances \(d = 75\)~mm and \(d = 90\)~mm are shown in Fig.~\ref{fig:false-color5} (a-c). The digit "2" is the object closer to the detector, at a distance of 75~mm, while the digit "1" is further away, at a distance of 90~mm. From the reconstruction results, it can be seen that the reconstruction of digit "2" is clearer compared to digit "1". This is because more information from digit "2" is captured by the detector. However, for digit "1" which is farther from the detector, the reconstruction result is less satisfactory. In scenarios (a2) and (c2), where the overlap between the two objects is minimal, the reconstruction is still acceptable. Although there is some shadowing from digit "2", the reconstruction of digit "1" does not exhibit deformation or truncation. In scenario (b2), however, where the overlap is more significant, digit "1" shows discontinuities and deformations, indicating a substantial impact.

Next, we applied our proposed method. Our method only requires the diffraction patterns as input, specifically the left-side images from Fig.~\ref{fig:false-color4} (c-e), without using the phase information. In other words, heterodyne detection is no longer necessary. 
The results obtained are shown in Fig.~\ref{fig:false-color5} (d-f). It can be observed that for the reconstruction of digit "2", even in the scenario with the heaviest overlap (e1), a clear result free from ghosting is displayed. For the reconstruction of digit "1", certain parts of digit "2" form some influence, resulting in shadows that do not belong to digit "1" in Fig.~\ref{fig:false-color5} (d2-f2). However, compared to (a2-c2), there is a significant improvement. Specifically, in scenarios (d2) and (f2), the outline and shape of digit "1" are more complete and accurate, with significantly reduced ghosting and deformation. Even in scenario (e2), where the overlap is most severe, a relatively clear reconstruction of digit "1" can still be observed.
In summary, our proposed method significantly improves the quality of reconstructed images without using phase information. It effectively reduces noise and deformation even in cases with severe ghosting and object overlap, ensuring the clarity and accuracy of the images. 

\begin{figure}[ht]
\centering
\includegraphics[width=\linewidth]{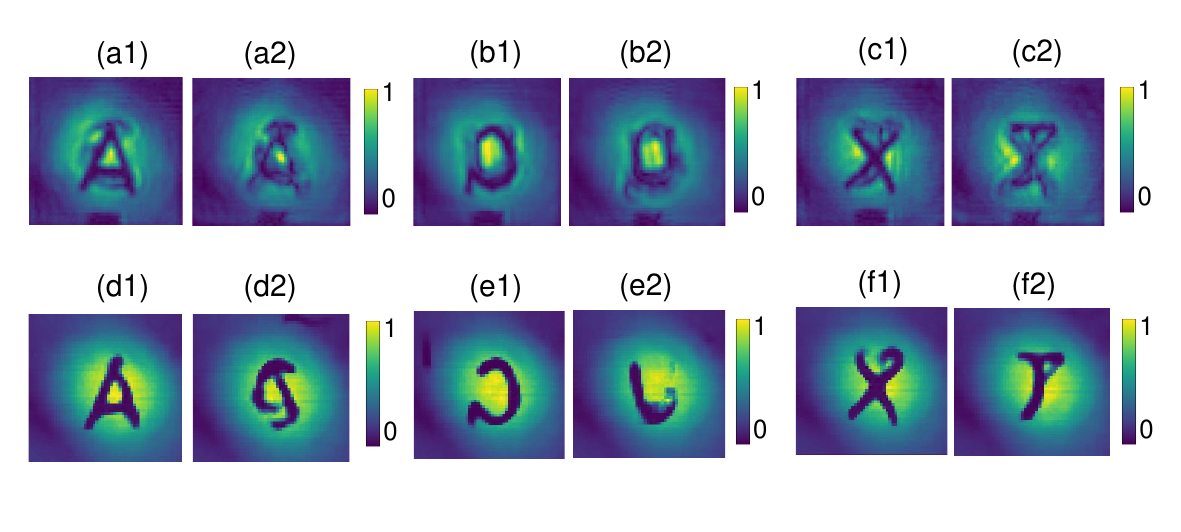}
\caption{Reconstruction results of more objects using the back-propagation method and our proposed self-training neural network.}
\label{fig:false-color6}
\end{figure}

To validate the generalization capability of our method, we reconstructed objects different from the MNIST dataset, as shown in Figure~\ref{fig:false-color6}. These objects are combinations of the letters 'AC', 'GU', and 'XJ'. Similarly, we first used both amplitude and phase for back-propagation, and the results are shown in Fig.~\ref{fig:false-color6} (a-c). When the overlap between the objects is significant, the reconstruction of the second object ('A', 'G', 'X') is slightly better than that of the first object ('C', 'U', 'J'), but both are heavily disrupted and cannot be independently reconstructed.

Next, we used only a single diffraction pattern as input to our neural network, and the prediction results are shown in Fig.~\ref{fig:false-color6} (d-f). It can be seen that the predictions for the objects closer to the detector are quite accurate, especially for the letter 'G', which is almost completely reconstructed without the influence of the other object. For the objects farther from the detector, although the influence of the second object cannot be entirely eliminated, the results are still better than those obtained by back-propagation, despite using less input information.
Therefore, our method demonstrates good generalization performance on different datasets, especially in complex scenarios where it effectively separates and reconstructs overlapping objects. By using only diffraction patterns as input, our method reduces computational complexity while still providing high-quality reconstruction results.

\section{Conclusion}
In conclusion, we propose a novel method for reconstructing occluded objects in THz holographic systems using a physics-based self-training algorithm, validated with experimental data. Initially, we pre-train the supervised learning network using only simulated data. Subsequently, we integrate a small amount of collected experimental data with the simulated data for further training. The self-training process is crucial, as illustrated in \nameref{AppendixB}, where we compare it to using supervised learning alone.
This method significantly outperforms traditional optical reconstruction methods, such as back-propagation, by effectively eliminating mutual interference between two objects. Another major advantage is that it no longer requires phase information; a single diffraction pattern input is sufficient for prediction. This implies that power detection could potentially replace heterodyne detection to some extent. Moreover, by using simulated data exclusively for pre-training, we address the challenge of data acquisition in the THz range.

\bibliography{sample}

\section*{Appendix A}
\label{AppendixA}

\begin{figure}[ht]
\centering
\includegraphics[width=\linewidth]{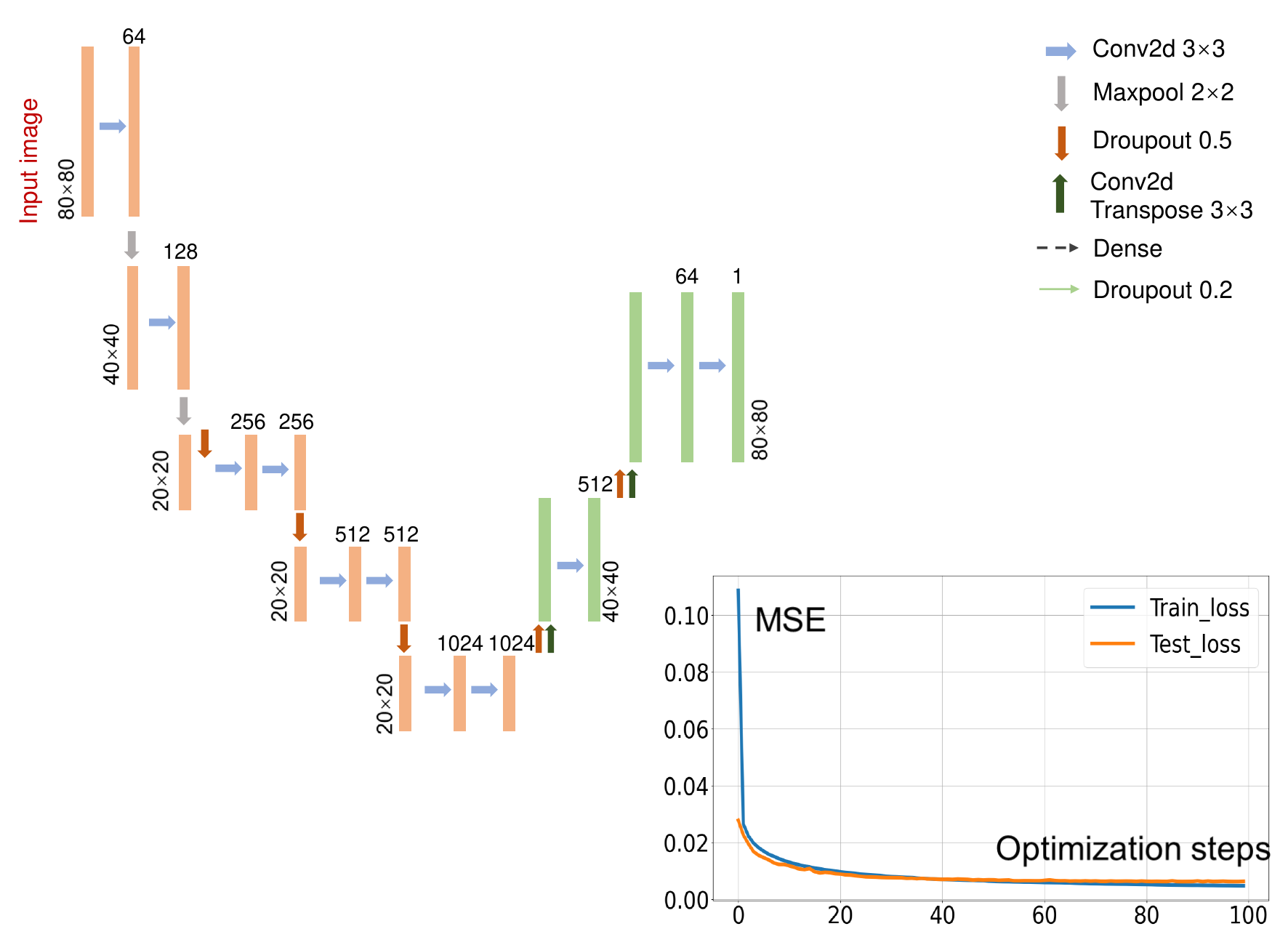}
\caption{Structure of the proposed NN and training curve}
\label{fig:false-color7}
\end{figure}

Figure~\ref{fig:false-color7} illustrates the architecture of our neural network. During the pre-training phase, we used the MNIST dataset to generate training and testing data, consisting of 30,000 training samples and 3,000 testing samples. The proposed algorithm's architecture was implemented using TensorFlow, an open-source deep learning framework \cite{abadi2016tensorflow}. We employed the Adam optimizer \cite{kingma2014adam} with a learning rate of 0.0001 to optimize the network's weights and biases. To achieve better convergence, we added noise uniformly distributed between 0 and $\frac{1}{30}$ to the fixed input diffraction patterns during each optimization step.
As shown in Figure~\ref{fig:false-color7}, the encoding step (downsampling part) includes convolutional layers with a kernel size of $3\times 3$ (blue arrows) with ReLU activation functions and max pooling layers with a kernel size of $2\times 2$ (gray arrows). The decoding process (upsampling part) consists of transposed convolutional layers with a kernel size of $3\times 3$ (dark green arrows) and additional convolutional layers. The input images have a size of $80\times 80$ pixels. The network was trained for 1000 epochs, during which the loss for both the training and testing sets consistently decreased until the MSE for the training set reached 0.0035, and the testing set's MSE stabilized around 0.0038. This indicates that the network did not overfit and that the training was successful.
The pre-training process took approximately 2 hours on a PC equipped with a 16-core 3.50 GHz CPU, 64 GB RAM, and an Nvidia GeForce RTX 3080 GPU. The subsequent self-training process, following pre-training, used 1066 data samples with 10\% designated as the test set. This process took only $\sim \!\! 10$~ minutes.

\section*{Appendix B}
\label{AppendixB}

\begin{figure}[ht]
\centering
\includegraphics[width=\linewidth]{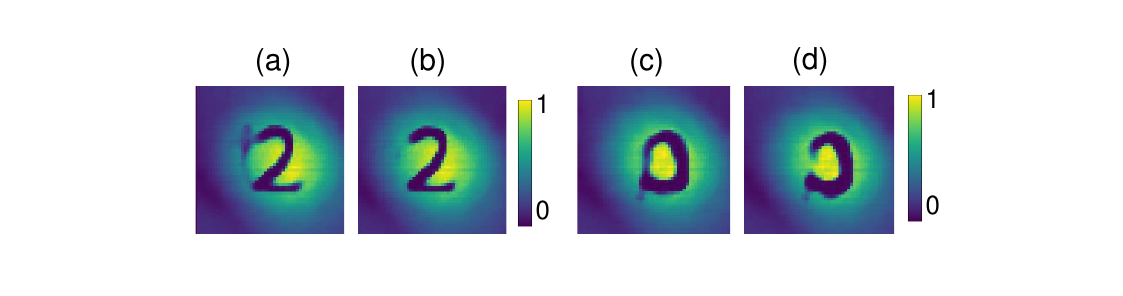}
\caption{Comparison between the reconstruction results of supervised NN and our self-training NN}
\label{fig:false-color8}
\end{figure}

Figure~\ref{fig:false-color8} presents a comparison between the reconstruction results of our self-training algorithm, which incorporates experimental data, and those obtained using a purely supervised learning network with the same amount of data. Fig.~\ref{fig:false-color8} (a) and (c) show the prediction results from the supervised learning network, while (b) and (d) display the predictions from our proposed method. Regardless of whether the predictions involve digits or letters, our method demonstrates the advantage of minimizing the influence of another object. For instance, in Fig.~\ref{fig:false-color8} (a), the left edge of digit '2' still retains the shadow of digit '1', and in Fig.~\ref{fig:false-color8} (c), the letter 'G' is still affected by the letter 'U'. Therefore, the advantages of our method are clearly evident.

\end{document}